\documentclass[12pt]{article}
\usepackage[dvips]{graphicx}
\def\imo{i}
\def\email{}

\begin{document}\sloppy
\begin{center}
{\Large\bf Quasi-normal modes of the scalar hairy black hole}\\
\vspace{0.8cm}
{\bf\tt Alexander Zhidenko}\\
Department of Physics, Dniepropetrovsk National University\\
St. Naukova 13, Dniepropetrovsk 49050, Ukraine\\
\email{Z\_A\_V@ukr.net}
\bigskip \nopagebreak
\end{center}

\begin{abstract}
We calculate QNMs of the scalar hairy black hole in the AdS background using
Horowitz-Hubeny method for the potential that is not known in analytical form. For some
black hole parameters we found pure imaginary frequencies. Increasing of the scalar field
mass does not cause the imaginary part to vanish, it reaches some minimum and then
increases, thus in the case under consideration the infinitely long living modes
(quasi-resonances) do not appear.
\end{abstract}

\section{Introduction}
There are three stages of the perturbation evolution near a black hole. The initial black
hole outburst from the perturbation source takes place at early times of the evolution.
Then at late times the damping (\textit{quasi-normal}) oscillations appear, finally
suppressed by power-law or exponential tails at asymptotically late times. The
quasi-normal modes (QNMs) can be characterized by a set of complex frequencies: real part
is the actual oscillation frequency while imaginary part describe the damping rate of the
oscillation. The QNMs do not depend on initial perturbations being, thereby, an important
characteristic of the black hole.

For this reason, QNMs of black holes in asymptotically flat space-time have been
extensively studied for more then thirty years (see reviews \cite{QNMrev}). Yet the
quasi-normal spectrum of a black hole is crucially dependent on a cosmological background
which a black hole is immersed in. QNMs of black holes in de Sitter background were
considered recent years in
\cite{gen,dS}.

Despite we live in the de Sitter universe, black holes in the
\mbox{\textit{anti-de Sitter}(AdS)} background attracted
considerable interest due to the famous \mbox{AdS/CFT} correspondence
\cite{Maldacena.1998}: it is shown that large black holes in AdS space-time correspond to
approximately thermal states in the dual \mbox{\textit{conformal field theory}(CFT)}.
According to this correspondence, QNMs of a black hole should coincide with poles of
retarded Green function of the perturbation in ${\cal N}=4$ $SU(N)$ super-Yang-Mills
theory for large $N$ and were extensively studied recent years
\cite{gen,AdS}.

First of all, let us briefly review what is known about perturbations of a scalar field
interacting with a black hole. Scalar field is the simplest model, frequently used,  when
we neglect spin influence \cite{Leaver, IyerWill}. Also scalar field is used to explain
dark matter problem (see e. g. \cite{SeidelSuen}). In fact almost all papers which
considered scalar field perturbations and corresponding quasi-normal modes were limited
by minimal coupling of the scalar field with a black hole, i. e. by consideration of test
scalar field propagating in the black hole background \cite{scalardS, scalarAdS}. Further
extensions within minimal coupling were made to consider massive scalar field
\cite{masscalar,TamakiNomura}.

Very few papers beyond the minimal coupling are works where perturbations of a black hole
with specific form of scalar field called dilaton were considered
\cite{dilaton,TamakiNomura}, and also perturbations of
Bocharova-Bronnikov-Melnikov-Bekenstein (BBMB) black hole
\cite{BBMBBH} studied in \cite{KonoplyaMolina}. Note however, that BBMB black hole is
topologically equivalent to extremal Reissner-Nordstr\"om solution and therefore does not
provide any unexpected features
\cite{KonoplyaMolina}.

Since the paper of \mbox{J. Bekenstein} \cite{Bekenstein}, it is well-known that black
hole can not have scalar hair within minimal coupling. Yet as recently have been found in
\cite{Winstanley.2003,Winstanley.2005}, there is a possibility of dressing a
four-dimensional black hole in AdS space-time with a non-minimally coupled classical
scalar field. In \cite{RaduWinstanley} the results were extended for higher dimensional
configurations. Moreover, fortunately the found solution is stable, thereby allowing us
to study its quasi-normal modes. Thus one motivation to undergo this study is to find out
what will happen with the ringing behavior of a black hole with scalar hairs, Another aim
is rather technical: for the above black hole perturbations the problem is reduced to
wave-like equations but with effective potential which is not known in analytical form,
but expressed in terms of unknown functions obeying some differential equations.

This is the case of many physically relevant situations such as dirty black holes (i. e.
with some distribution of matter around them), black holes with massive dilaton,
evaporating black holes and other objects determined by a set of equations that can not
be solved analytically. We hope therefore that the way we are considering quasi-normal
spectrum here may be of some use in those cases.

The other interesting issue is related to arbitrary long living modes (in linear
approximation) which were called quasi-resonances. They appear when perturbing massive
scalar \cite{OhashiSakagami} or vector \cite{Konoplya} fields near a black hole of a
specific mass. In \cite{KonoplyaZhidenko} it was shown, that quasi-resonances are
permitted for black holes immersed in asymptotically flat background and are not
permitted for asymptotically de Sitter space-times. We show that anti-de Sitter boundary
conditions do not permit quasi-resonances too (see the appendix). Our numerical
investigations confirm this statement.

This paper is organized as follows. In section \ref{basic} we review briefly
\cite{Winstanley.2003,Winstanley.2005} to introduce the hairy black hole and the
perturbation equation near it. In section \ref{calc} we describe the method we used for
our calculations. And finally, in section \ref{results} we make some comments about the
results.

\section{Scalar hairy black hole}\label{basic}
We consider spherically symmetric solution
\begin{equation}
ds^{2} = N(r) e^{2\delta(r)} dt^2 - N(r)^{-1} dr^2 - r^2
\left( d\theta ^{2} +
\sin \theta ^{2} \, d\varphi ^{2} \right)
\label{metric}
\end{equation}
of the action, which describes a self-interacting scalar field
$\phi$ with non-minimal coupling to gravity:
\begin{equation}
S=\int d^{4}x \, {\sqrt {-g}} \left[
\frac {1}{2}\left( R -2\Lambda \right)
-\frac {1}{2} \left( \nabla \phi \right) ^{2} -\frac {1}{2} \xi R
\phi ^{2} -V(\phi ) \right] ,
\label{action}
\end{equation}
where $R$ is the Ricci scalar, $\Lambda $ is the cosmological constant, $\xi $ is the
coupling constant, $V(\phi )=\displaystyle\frac{\mu^2\phi^2}{2}$ is the massive scalar field self-interaction
potential and $\left(
\nabla
\phi
\right) ^{2} = \nabla _{\mu }
\phi
\nabla ^{\mu } \phi $.

In \cite{Winstanley.2005} stability of the non-trivial solution has been proven for
$0<\xi<3/16$, $1-\xi\phi^2>0$, $\mu=0$. In \cite{Winstanley.2003}
some arguments for its stability for $0\leq\mu^2\leq-4\xi\Lambda$ are given and
the stability is proven for the particular case of $\xi=1/6$. In the
present paper we consider the above range of the parameters and
our results support stability of the solution within linear perturbatuon approach. 

It is convenient for our purposes to use conformal transformation in
this
region to simplify the equations of motion:
\begin{equation}
{\bar {g}}_{\mu \nu } =(1-\xi
\phi ^{2}) g_{\mu \nu }.
\label{conf_trans}
\end{equation}

After the transformation the action takes the form:
\begin{equation}
S=\int d^{4}{\bar x} \, {\sqrt {-{\bar {g}}}} \left(
\frac{1}{2} \left( {\bar {R}}-2\Lambda  \right)
-\frac {1}{2} \left( {\bar {\nabla }} \Phi \right) ^{2} -U(\Phi )
\right),
\label{conf_trans_action}
\end{equation}
where we define
\begin{equation}
\Phi =\int d\phi \sqrt{
\frac {(1-\xi \phi ^{2}) + 6\xi ^{2} \phi ^{2}}{
(1-\xi \phi ^{2})^{2}} },\qquad U(\Phi ) = \frac {V(\phi )+\Lambda
\xi \phi ^{2} \left( 2- \xi
\phi ^{2} \right) }{ \left( 1- \xi \phi ^{2} \right) ^{2}}.
\label{conf_trans_field}
\end{equation}

The metric takes the following form
\begin{equation}
d{\bar {s}}^{2} = {\bar {N}}({\bar {r}}) e^{2{\bar {\delta
}}({\bar {r}})} dt ^{2} - {\bar {N}}({\bar {r}}) ^{-1} d{\bar
{r}}^{2} - {\bar {r}}^{2} \left( d\theta ^{2} +
\sin \theta ^{2} \, d\varphi ^{2} \right),
\label{conf_trans_metric}
\end{equation}
where
\begin{eqnarray}\nonumber
{\bar {r}} &=& \left( 1 - \xi \phi ^{2} \right) ^{\frac {1}{2}}
r,\\\nonumber {\bar {N}} &=& N \left( 1- \xi \phi ^{2} -\xi r
\phi
\phi' \right) ^{2}
\left( 1- \xi \phi ^{2} \right) ^{-2},\\\nonumber
{\bar {N}} e^{\bar {2\delta }} & = & N e^{2\delta }
\left( 1- \xi \phi ^{2} \right).
\end{eqnarray}

Varying the action one can find the equations \cite{Winstanley.2005}:
\begin{eqnarray}
\frac{d({\bar r}\bar N)}{d{\bar r}} & = & 1-\Lambda{\bar r}^2-{\bar
r}^2\left(\frac {\bar N}{2}
\left(\frac {d\Phi }{d{\bar {r}}}\right)^2
+ U(\Phi)\right),
\label{minE1}
\\
\frac {d{\bar {\delta }}}{d{\bar {r}}} & = &
\frac {{\bar {r}}}{2} \left( \frac {d\Phi }{d{\bar {r}}} \right)^{2},
\label{minE2}
\\
0 & = & {\bar {N}} \frac {d^{2}\Phi }{d{\bar {r}}^{2}} + \left(
{\bar {N}} \frac {d{\bar {\delta }}}{d{\bar {r}}} + \frac {d{\bar
{N}}}{d{\bar {r}}} + \frac {2{\bar {N}}}{{\bar {r}}} \right)
\frac {d\Phi }{d{\bar {r}}}
- \frac {dU}{d\Phi}.
\label{mincscal}
\end{eqnarray}

In order to obtain the explicit form of (\ref{conf_trans_metric}) we have to solve the
equations with some boundary conditions that are associated with the black hole
parameters. We are using the following parameters. The event horizon radius ${\bar r}_h$
can be chosen arbitrary in order to fix length scale and defines boundary condition for
$\bar N$: ${\bar N}({\bar r}_h)=0$. We measure all dimensional units in units of ${\bar
r}_h$, thus we choose ${\bar r}_h=1$. ${\bar\delta}({\bar r}_h)$ can be also chosen
arbitrary to fix time scale. In order to introduce the boundary conditions in the same
point we choose ${\bar\delta}({\bar r}_h)=0$ as distinct from Winstanley's
${\bar\delta}(\infty)=0$. And the last parameter $\phi({\bar r}_h)=Q$ can be associated
with the scalar charge of the black hole.

Spherically symmetric perturbation equation for $\exp(-\imo\omega t)\psi(r) =r\delta\Phi$
takes the Schr\"{o}dinger wave-like form:
\begin{equation}
\frac {d
^{2} \psi }{d {\bar r}^{*2}} + (\omega^2-{\cal {U}})
\psi = 0,
\label{perturbed}
\end{equation}
where the perturbation potential ${\cal {U}}$ is given by
\cite{Winstanley.2003}:
\begin{equation}
{\cal {U}}  =
\frac {{\bar {N}}e^{2{\bar {\delta }}}}{{\bar {r}}^{2}}
\left(
(1  - (U+\Lambda ){\bar r}^2)\left(1-{\bar r}^2\left( \frac {d\Phi
}{d{\bar {r}}} \right) ^{2}\right) - {\bar N} +2{\bar {r}}^{3}
\frac {dU}{d\Phi }
\frac {d\Phi }{d{\bar {r}}}
+ {\bar {r}}^{2} \frac {d^{2}U}{d\Phi ^{2}}\right).
\label{perturb_potential}
\end{equation}
${\bar r}^*$ is tortoise coordinate given by
\begin{equation}
\frac {d{\bar r}^*}{d{\bar {r}}} =
e^{\bar\delta}{\bar N} ,
\label{tortoise}
\end{equation}
and similar to ordinal anti-de Sitter case:
$$\begin{array}{rcl}
{\bar r}={\bar r}_h&\quad\Longleftrightarrow\quad&{\bar r}^*=\infty,\\
{\bar r}=\infty&\quad\Longleftrightarrow\quad&{\bar r}^*=0.
\end{array}$$

This perturbation corresponds to infinitesimal changing of the black hole mass and is
similar to the zero-multipole scalar perturbations near a hairless black hole. The
crucial difference is that the perturbation of a scalar field due to the hair gives a
first-order correction to the metric.

\section{QNMs calculation}\label{calc}
Quasi-normal modes can be defined as eigenvalues of (\ref{perturbed}) with the boundary
conditions which require ingoing wave at the event horizon and zero at spatial infinity.
These boundary conditions are natural and mean that the black hole does not emit
radiation and our solution is not divergent. These specific boundary conditions allow us
to fix the one of the two independent solutions in both points and calculate eigenvalues
by comparing series near each of them. This technique is the base for two main precise
methods of numerical calculation of QNMs: the continued fraction method
\cite{Leaver} and the Horowitz-Hubeny method
\cite{Horowitz-Hubeny}. They both were extensively used to find
eigenfrequencies of perturbation equations with potential given analytically as
polynomial fraction. Fortunately Horowitz-Hubeny method is possible to be adopted for
equations with unknown potential.

Indeed if the potential ${\cal {U}}$ would be polynomial fraction we could eliminate
singularity at the horizon introducing new function
\begin{equation}\label{yexp}
y(z)=z^{\imo\omega\kappa}\psi(z), \qquad z=1-\frac{{\bar r}_h}{\bar r},
\end{equation}
where $\kappa$ is chosen to make $y(z)$ regular at the horizon $z=0$. The requirement for
$y(z)$ to be regular at the horizon fixes that independent solution which satisfies the
horizon boundary condition. To use the other boundary condition one needs to find an
equation for $y(z)$ in the form:
\begin{equation}\label{exp_eq}
s(z)z^2\phi''(z)+t(z)z\phi'(z)+u(z)\phi(z)=0,
\end{equation}
where $s(z)$, $t(z)$ and $u(z)$ are polynomial functions. Using
explicit form of the polynoms:
$$
s(z)=\sum_{n=0}^{M_s}s_nz^n, \qquad t(z)=\sum_{n=0}^{M_t}t_nz^n,
\qquad u(z)=\sum_{n=1}^{M_u}u_nz^n,
$$
it is easy to find series expansion for $\displaystyle y(z)=1+\sum_n y_nz^n$ (we suppose
that there is no singularities in the circle $|z|<1$, if singularities appear one needs
to use another definition of $z$ to avoid them):

\begin{equation}\label{yserie}
y_n=-\sum_{k=0}^{n-1}\frac{y_k(k(k-1)s_{n-k}+kt_{n-k}+u_{n-k})}{n(n-1)s_0+nt_0},
\end{equation}
and find $\omega$ by minimizing $|y(1)|$ restricted by some large number $N$ of expansion
terms in (\ref{yserie}).

It is clear that if $s(z)$, $t(z)$ and $u(z)$ are series and $s(1)$, $t(1)$ and $u(1)$
converge quickly enough $y_n$ is still possible to calculate because it depends
insignificantly on higher terms of the series.

Thus we use the following procedure of the QNMs calculation. We introduce $g({\bar
r})={\bar N}({\bar r})/{\bar r}^2$ that is finite in the region outside black hole ${\bar
r}_h<{\bar r}<\infty$. Using (\ref{perturbed}) we express $s(z)$, $t(z)$ and $u(z)$ in
terms of $g(z)$, $\phi(z)$ and ${\bar\delta}(z)$ and find that $\kappa=1/g'(0)$.

From (\ref{minE1}, \ref{minE2}, \ref{mincscal}) we calculate for given black hole
parameters the first $M\gg1$ coefficients in the series expansions for $g(z)$, $\phi(z)$
and ${\bar\delta}(z)$. Then for each given $\omega$ we find the first $M$ coefficients of
$s(z)$, $t(z)$ and $u(z)$ and using (\ref{yserie}) the first $N\gg M$ terms of $y(z)$.
And, finally, we minimize $|y(1)|$ with respect to $\omega$.

We increase $M$ and $N$ for every given $M$ until the value of $\omega$ stops changing
within required precision. If we find finite $M$ and $N$ for a given precision of
$\omega$, we believe that the procedure converges and the $\omega$ is eigenvalue of
(\ref{perturbed}). To control convergence of the procedure we used the simplest
minimization method: we start from some initial value of $\omega$ and calculate $y(1)$
for this value, then we change $\omega$ by small value and calculate $y(1)$ for it doing
so while its modulus is decreasing, after that we decrease the small value and repeat the
procedure until the required precision is reached.

Obviously, the series $s(z)$, $t(z)$ and $u(z)$ are definite up to an arbitrary finite
function. Since we require quick convergence of them the selection of such function can
greatly decrease $M$ and $N$ and therefore the calculation time. We used $(1-z)^k$ factor
for this purpose. (The complete procedure of the QNMs searching written in MATHEMATICA is
available from the author upon request.)

\section{Discussion}\label{results}
As were noticed above the method is appropriate only if equation (\ref{exp_eq}) has no
singular points in the circle $|z|<1$. As long as functions $s(z)$, $t(z)$ and $u(z)$ are
known only as series we can not be sure that it is true within our approach. We believe
that it is true because our results are stable against small changing of any of free
parameters. If the equation had singularities in the unit circle we would obtain random
results that do not approximate a smooth curve. Unfortunately we have no results to
compare with ours because the hairy black hole is found as non-trivial solution of
non-minimally coupled scalar field to gravity and has no studied limit to compare.
Therefore calculation with an alternative method would be welcome to check the obtained
results.

\begin{figure}
\begin{center}
\includegraphics{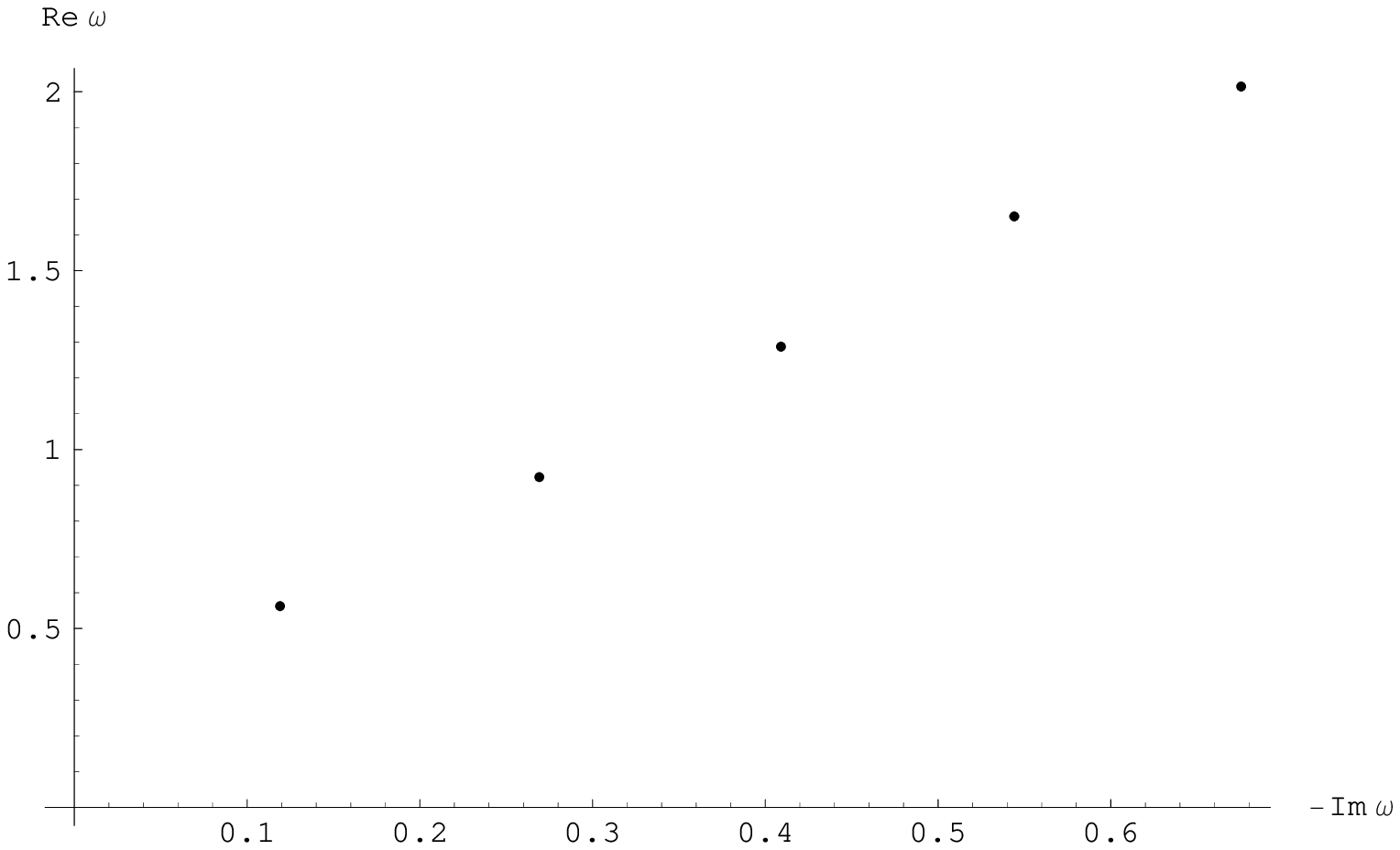}
\caption{\textbf{First five quasi-normal frequencies for $\Lambda=-0.1$,
  $Q=1$, $\xi=0.1$, $\mu=0$.} They approach equidistant spacing as well
  as ordinary blach hole ones in AdS space-time.}\label{Nfig}
\end{center}
\end{figure}

Taking into account these argues we restricted our calculations to fundamental
modes\footnote{Fundamental QNM is the eigenfrequency with the lowest absolute value of
its imaginary part.} though the method allows to find higher ones that are presented for
particular values of the parameters (see figure \ref{Nfig}). One can see that higher
overtones tend to equidistant spacing similar to hairless black hole spectrum behavior in
AdS \cite{CardosoKonoplyaLemos}. Yet fundamental modes give us a qualitative picture of
quasi-normal spectrum depending on the free parameters: $\xi$, $Q$, $\Lambda$ and $\mu$
(we measure all quantities in ${\bar r}_h$ units).

\begin{figure}
\begin{center}
\includegraphics{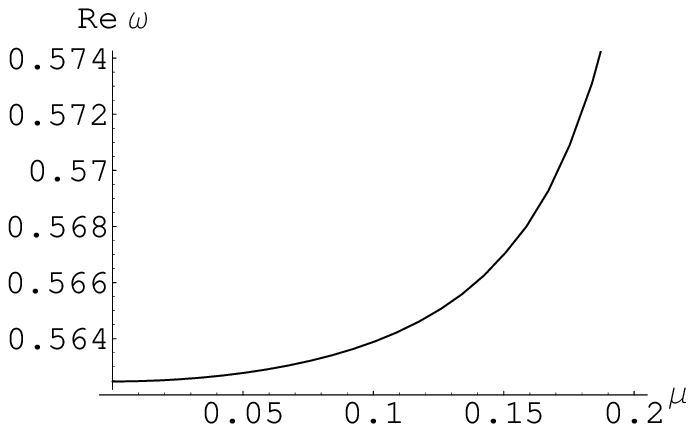}\includegraphics{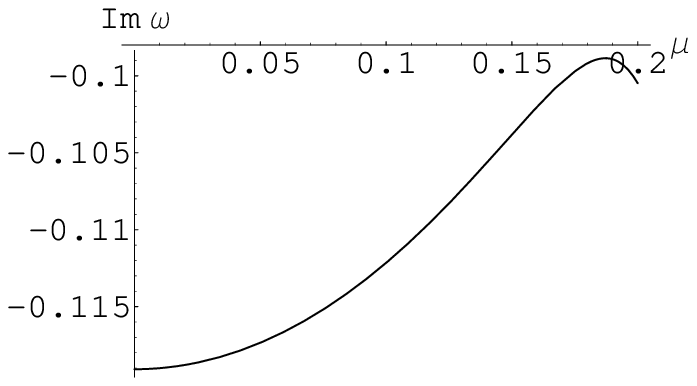}\\($\Lambda=-0.1$, $Q=1$, $\xi=0.1$)
\includegraphics{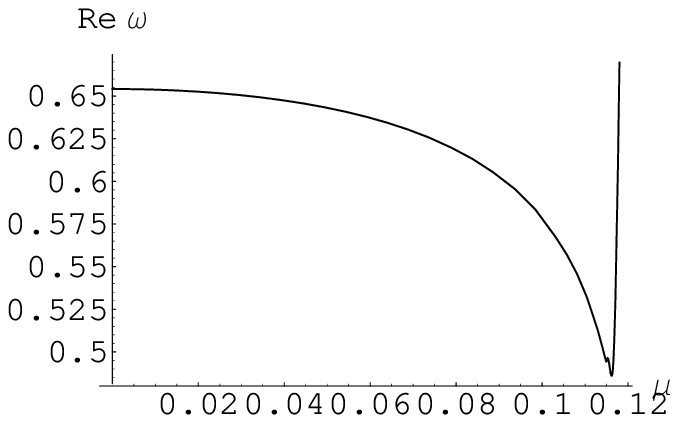}\includegraphics{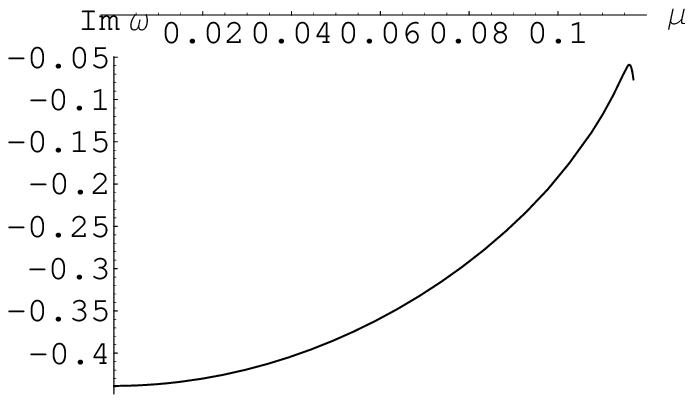}\\($\Lambda=-0.05$, $Q=2.2$, $\xi=0.1$)
\caption{\textbf{Dependance of the real and imaginary part of QN frequency  on $\mu$.}
We consider only $0\leq\mu^2\leq-4\xi\Lambda$.  The real part changes
within comparatively small range.
The imaginary part reaches its maximum remaining negative. For this
value of $\mu$ the oscillations have the longest lifetime, but they still
damp and we do not observe quasi-resonanses. For larger scalar field
masses the imaginary part of frequency decreases and the damping rate is higher.
}\label{mufig}
\end{center}
\end{figure}

Imaginary part demonstrates the most predictable behavior. It changes monotonously with
respect to $\Lambda$, $Q$ and $\xi$. Scalar field mass cause imaginary part to increase
but not to vanish. It reaches a maximum for some particular mass and then decrease very
rapidly (see figure
\ref{mufig}). Thus no infinitely long living oscillations appear. This fact is not
unexpected and seems to be general for all black holes in the asymptotically anti-de
Sitter background.

\begin{figure}
\begin{center}
\includegraphics{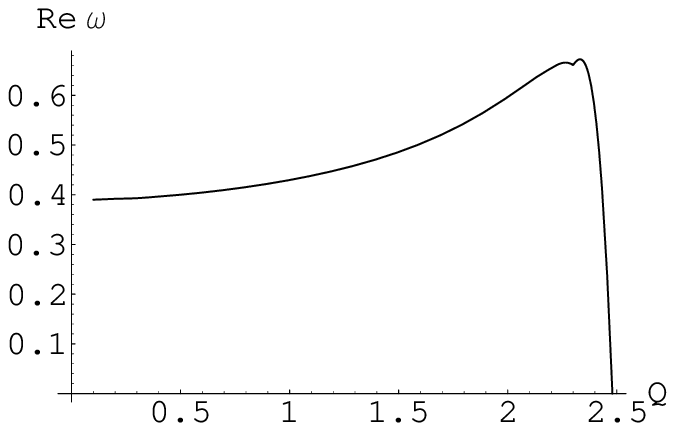}\includegraphics{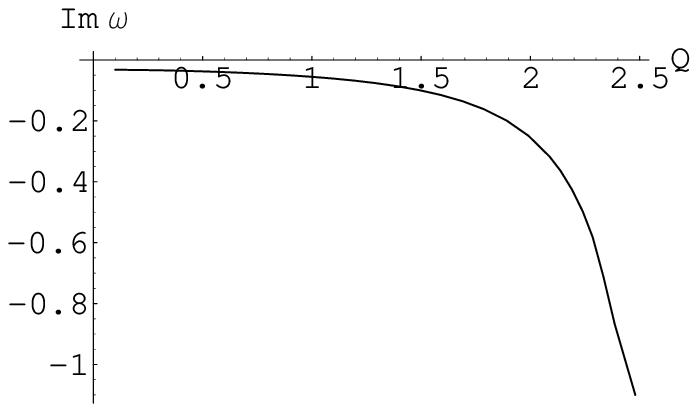}\\($\Lambda=-0.05$, $\xi=0.1$, $\mu=0$)
\includegraphics{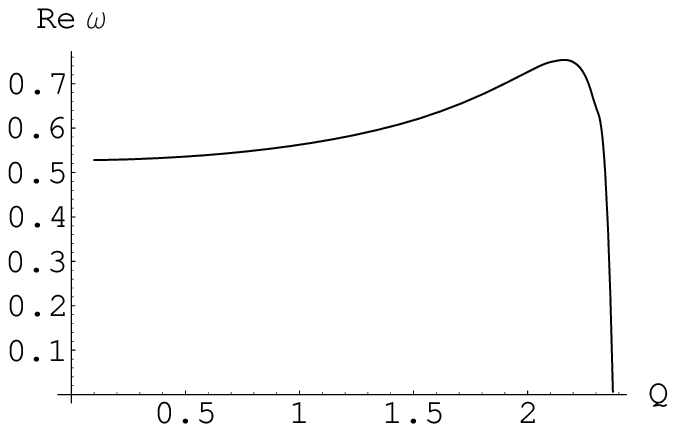}\includegraphics{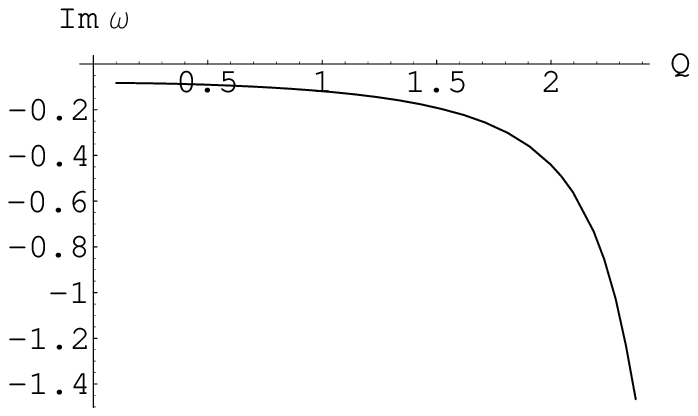}\\($\Lambda=-0.1$, $\xi=0.1$, $\mu=0$)
\caption{\textbf{Dependance of the real and imaginary part of QN frequency on $Q$.}
The imaginary part quickly decreases with increasing of the ``scalar
charge'' $Q$. The real part reaches its maximum and then quickly falls
down to zero. It vanishes for some non-critical charge
$Q_0<\xi^{-1/2}$ that depends on other parameters: $\Lambda$, $\xi$, $\mu$. For $Q\geq Q_0$ the real part remains zero and the
frequency is purely imaginary (within numerical precision).
}\label{Qfig}
\end{center}
\end{figure}

\begin{figure}
\begin{center}
\includegraphics{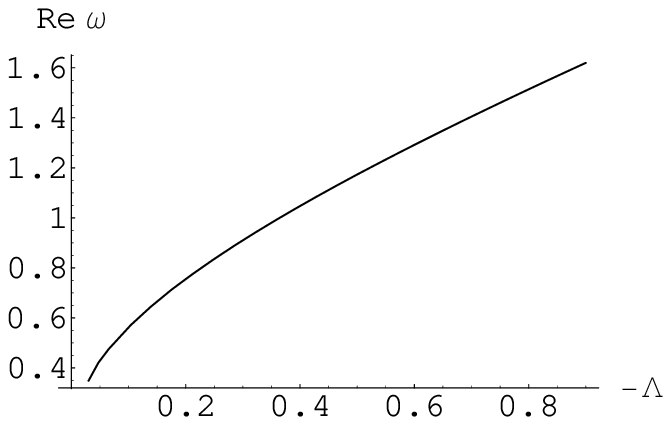}\includegraphics{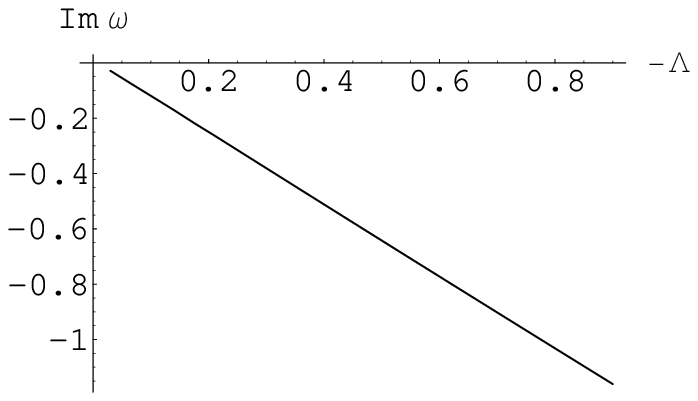}\\($Q=1$, $\xi=0.1$, $\mu=0$)
\includegraphics{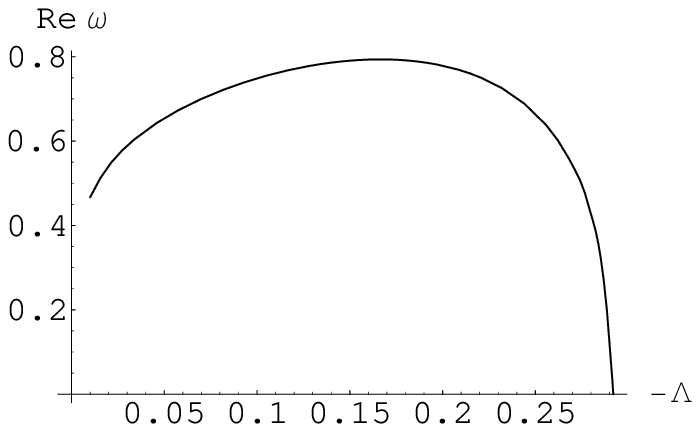}\includegraphics{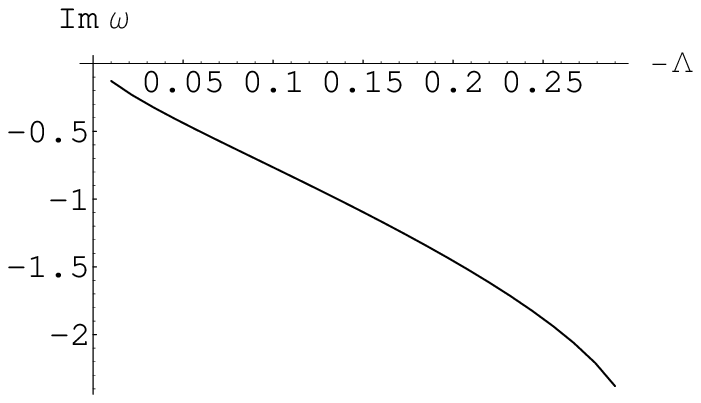}\\($Q=2.2$, $\xi=0.1$, $\mu=0$)
\caption{\textbf{Dependance of the real and imaginary part of QN frequency on  $\Lambda$.}
The real part increases for relatively small ``scalar charges'' $Q$, but for large
$Q=2.2$ it also reaches maximum and then falls down and vanishes $\Lambda\approx-0.292$. For larger values of $\Lambda$ the
real part remains zero and purely imaginary frequencies exist.
The imaginary part decreases almost linearly with $\Lambda$. You can
see a small deflection from linearity on the bottom graphic. This
deflection appears in the region where the real part decreases.
}\label{Lfig}
\end{center}
\end{figure}

\begin{figure}
\begin{center}
\includegraphics{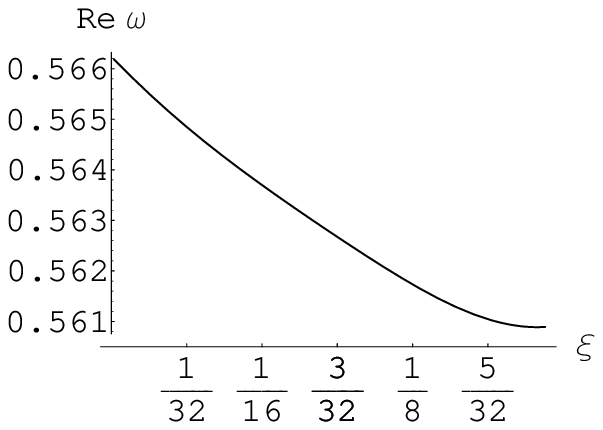}\includegraphics{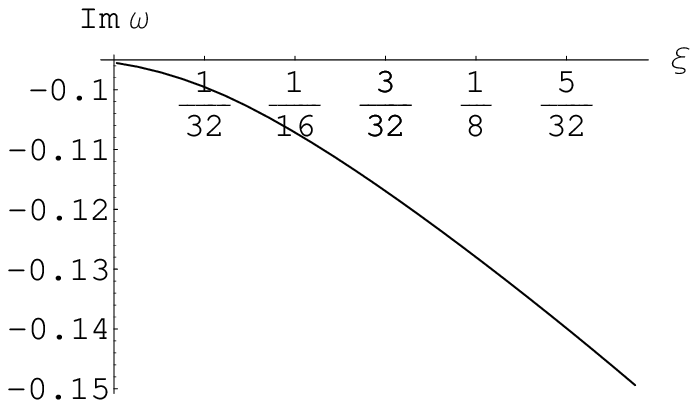}\\($Q=1$, $\Lambda=-0.1$, $\mu=0$)
\includegraphics{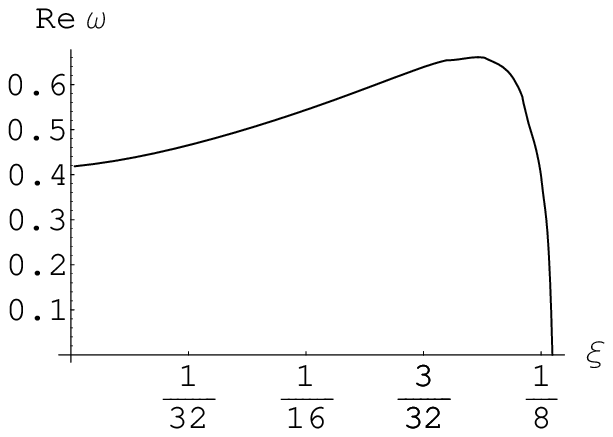}\includegraphics{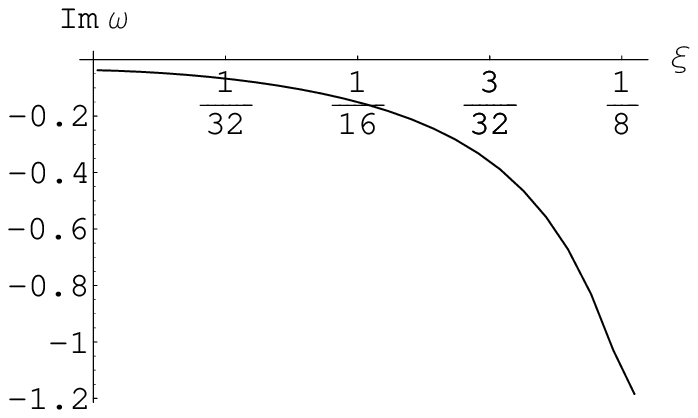}\\($Q=2.2$, $\Lambda=-0.05$, $\mu=0$)
\caption{\textbf{Dependance of the real and imaginary part of QN frequency on $\xi<\frac{3}{16}$.}
Dependance of the quasi-normal frequency on $\xi$ is different for
different values of $\Lambda$ and $Q$. The imaginary part decreases
with increasing of $\xi$. The quickness of such decreasing depends on
$\Lambda$ and $Q$. For some $\Lambda$ and $Q$ (as presented on the
bottom graphic) the real part can fall down
to zero and purely imaginary frequencies appear for larger $\xi$.
}\label{xifig}
\end{center}
\end{figure}

Real part has more complicated behavior and depends significantly on the parameters $Q$
and $\Lambda$ (see figures
\ref{Qfig} and \ref{Lfig}). Other parameters
$\xi$ and $\mu$ change the real part within the comparable small bounds (figures
\ref{xifig} and \ref{mufig}). For relatively large values of $\xi$,
$\Lambda$, $Q$ purely imaginary frequencies appear.

Also we did not find any growing modes (with positive imaginary part) that supports
stability of such hairy black holes.

\section*{Appendix: No quasi-resonance existence proof}
In \cite{KonoplyaZhidenko} was shown that the existence of purely real quasi-normal modes
(quasi-resonances) of massive scalar field near Schwarzschild black hole is due to
specific boundary conditions at spatial infinity for the case of asymptotically flat
space-time. It was explained also why there is no quasi-resonances of massive scalar
field in the asymptotically de Sitter background if the potential is zero at both the
horizons. Though the potential kind and the boundary conditions for asymptotically
anti-de Sitter space-time are different one can show similarly that no quasi-resonances
can appear in this background too.

Indeed, suppose $\omega$ is real and consider an imaginary part of the integral of
(\ref{perturbed}):
\begin{eqnarray}\nonumber
&&\intop_{{\bar r}^*=-\infty}^{{\bar r}^*=0}\psi^*\left(\frac {d
^{2} \psi }{d {\bar r}^{*2}} + (\omega^2-{\cal {U}})
\psi\right)d{\bar r}^* = \\ &&=\psi^*\frac{d\psi}{d{\bar r}^*}\Biggr|_{{\bar {r}}^*=-\infty}^{{\bar {r}}^*=0}+
\intop_{{\bar r}^*=-\infty}^{{\bar r}^*=0}\left(-\left|\frac {d
\psi }{d {\bar r}^*}\right|^2 + (\omega^2-{\cal {U}})
|\psi|^2\right)d{\bar r}^*= 0.
\end{eqnarray}
Because $\cal {U}$ is real
\def\Im#1{\mbox{Im}\left(#1\right)}
\begin{eqnarray}0&=&\Im{\psi^*\frac{d\psi}{d{\bar r}^*}\Biggr|_{{\bar {r}}^*-\infty}^{{\bar {r}}^*=0}+
\intop_{{\bar r}^*=-\infty}^{{\bar r}^*=0}\left(-\left|\frac {d
\psi }{d {\bar r}^*}\right|^2 + (\omega^2-{\cal {U}})
|\psi|^2\right)d{\bar
r}^*}=\\\nonumber&&=\Im{\psi^*\frac{d\psi}{d{\bar
r}^*}\Biggr|_{{\bar r}^*=-\infty}^{{\bar r}^*=0}}.
\end{eqnarray}

Since we require Dirichlet boundary conditions at spatial infinity and
by definition $\displaystyle \lim_{{\bar
r}\rightarrow\infty}\frac{e^{\bar\delta}\bar N}{{\bar
r}^2}=-\Lambda_{eff}$,
$$\psi^*\frac{d\psi}{d{\bar r}^*}\Biggr|_{{\bar r}^*=0}=\psi^*e^{\bar\delta}{\bar N}\frac{d\psi}{d{\bar r}}\Biggr|_{{\bar r}=\infty}$$
is finite. Indeed convergence of (\ref{yexp}) at
spatial infinity implies that
\begin{equation}
\psi = z^{-\imo\omega\kappa}y(z) = \left(1-\frac{{\bar
r}_h}{\bar r}\right)^{-\imo\omega\kappa}\sum_{n=0}^\infty y_n
\left(1-\frac{{\bar r}_h}{\bar r}\right)^n=\sum_{n=1}^\infty C_n\left(\frac{{\bar r}_h}{\bar r}\right)^n,
\end{equation}
where $C_n$ can be expressed in terms of $y_n$ and thus
$\displaystyle {e^{\bar\delta}\bar N}\frac{d\psi}{d{\bar
r}}\Biggr|_{{\bar r}=\infty}=-C_1\Lambda_{eff}$.

Therefore
\begin{equation}
\Im{\psi^*\frac{d\psi}{d{\bar
r}^*}\Biggr|_{{\bar r}^*=-\infty}^{{\bar
r}^*=0}}=-\Im{\psi^*\frac{d\psi}{d{\bar r}^*}\Biggr|_{{\bar
r}^*=-\infty}}=\omega|y_0|^2=0,
\end{equation}
that is impossible because $y_0=0$ leads to trivial solution. Thus we prove that there
can be no quasi-resonances.

\section*{Acknowledgements}
I would like to thank R. Konoplya for proposing this problem to me and critical reading
of the manuscript.


\end{document}